\documentclass[12pt]{article}
\usepackage{amssymb,amsfonts}
\def \la{\lambda}

\def \a{\alpha}
\def \b{\beta}

\newcounter{map}
\newcounter{fel}
\newcounter{aw}
\newcounter{aww}
\begin{document}
\baselineskip 18pt

\title{How to solve Yang-Baxter equation using the Taylor expansion of 
$R$-matrix}
\author{P.~N.~Bibikov\thanks{bibikov@PB7855.spb.edu}}

\maketitle

\vskip5mm

\begin{abstract}
A new method for solving the Yang-Baxter equation is presented.
It is based on the Taylor expansion of $R$-matrix which is developed
up to the power $\la^6$. Using this method the $R$-matrix for integrable 
spin-ladder is calculated.
\end{abstract}

\begin{section}
{The power expansion formulas}
\end{section}
By the Yang-Baxter equation (in the braid group form) is called the following
equation \cite{1}:
\begin{equation}
R_{12}(\la-\mu)R_{23}(\la)R_{12}({\mu})=R_{23}(\mu)R_{12}({\la})R_{23}(\la-\mu),
\end{equation}
on the $N^2\times N^2$ matrix $R(\la)$ called $R$-matrix. Here for an 
arbitrary $N^2\times N^2$ matrix $X$ the $N^3\times N^3$ matrices
$X_{12}$ and $X_{23}$ are defined as 
$X_{12}=X\otimes I_N$, $X_{23}=I_N\otimes X$,
where $I_N$ is a unit $N\times N$ matrix.

In this paper we study only the regular solutions of the Eq. (1).
These solutions play an important role in the Quantum Inverse Scattering
Method (QISM) \cite{2} and additionally satisfy the following condition:
\begin{equation}
R(0)=I_{N^2},
\end{equation}

The system (1),(2) is invariant under multiplication of $R(\la)$ on
the arbitrary regular function $f(\la)$ satisfying the initial condition
$f(0)=1$:
\begin{equation}
R(\la)\rightarrow f(\la)R(\la).
\end{equation}
 
Our method of solving the system (1)-(2) is based on the Taylor expansion
of $R$-matrix:
\begin{equation}
R(\la)=I+\lambda H+\sum_{n=2}^{\infty}\frac{1}{n!}R^{(n)}\la^n.
\end{equation}
(From now we shall write $I$ instead of
$I_{N^2}$). The matrix $H$ according to the QISM plays a role of
the local Hamiltonian density
of the integrable model corresponding to $R(\la)$ \cite{2}.

Substituting the expansion (4) into the Eq. (1) we obtain the following
formulas for $R^{(n)}$, $n=2,...,6$:
\begin{eqnarray}
R^{(2)}&=&H^2,\\
R^{(3)}&=&H^3+K,\\
R^{(4)}&=&H^4+2(HK+KH),\\
R^{(5)}&=&H^5+L+2(KH^2+H^2K)+6HKH,\\
R^{(6)}&=&H^6+KH^3+H^3K+9(H^2KH+HKH^2)+\nonumber\\
&&\qquad\qquad\qquad\qquad\qquad+10K^2+3(HL+LH).
\end{eqnarray}
The $N^2\times N^2$
matrices $K$, $L$ appearing in (6)-(9) and the matrix $H$ satisfy a system 
of equations that follows from the system (1)-(2). The first equation: 
\begin{equation}
[(H_{12}+H_{23}),[H_{12},H_{23}]]=K_{23}-K_{12},
\end{equation}
is called the Reshetikhin condition \cite{3}. As a first integrability test for
the matrix $H$ it was studied in \cite{4},\cite{5}. 

The second equation:
\begin{eqnarray}
L_{23}-L_{12}&=&[H_{12}^3+H_{23}^3+3(K_{12}+K_{23}),[H_{12},H_{23}]]+\nonumber\\
&&+3(H_{12}[[H_{12},H_{23}],H_{12}]H_{12}+
H_{23}[[H_{12},H_{23}],H_{23}]H_{23})+\nonumber\\
&&(H_{12}H_{23}+H_{23}H_{12})(K_{23}-K_{12})+(K_{23}-K_{12})(H_{12}H_{23}+
H_{23}H_{12})-\nonumber\\
&&-2(H_{12}(K_{23}-K_{12})H_{23}+H_{23}(K_{23}-K_{12})H_{12}).
\end{eqnarray}
obtained in \cite{5} may be used as a second integrability test. 

We also may write the third integrability condition in the following form:
\begin{eqnarray}
R_{23}^{(6)}-R_{12}^{(6)}+4(H_{12}R_{23}^{(5)}-R_{12}^{(5)}H_{23})+
5(H_{12}^2R_{23}^{(4)}-R_{12}^{(4)}H_{23}^2)&=&\nonumber\\
H_{23}(R_{12}R_{23})^{(5)}-(R_{12}R_{23})^{(5)}H_{12}.&&
\end{eqnarray}

For almost all known $R$-matrices the quotients
of their matrix elements have a rather simple (rational, trigonometric or
elliptic) forms.
It was mentioned in the Ref. \cite{4} that using 
the first terms of power expansions for quotients of $R$-matrix elements
it may be possible to identify the full exact expressions for them. 
In the next section we show that this
approach really may be very effective and calculate a $4\times 4$ $R$-matrix 
for a spin-ladder model.

\section{The spin-ladder R-matrix}
The spin ladder models have attracted considerable attention in resent years 
due to the developing experimental results on ladder materials and the hope to
get insight into the physics of metal-oxide superconductors \cite{6}
In the Ref. \cite{7} the integrable spin-ladder system was presented. 
Its integrability was not proved rigorously but 
the model was solved via coordinate Bethe ansatz. The Hamiltonian studied in
the Ref. \cite{7} has the following density:
\begin{eqnarray}
H&=&\frac{\alpha}{4}
(\sigma_i\otimes I_2\otimes\sigma_i\otimes I_2+
I_2\otimes\sigma_i\otimes I_2\otimes\sigma_i-
\sigma_i\otimes I_4\otimes\sigma_i-\nonumber\\
&&I_2\otimes\sigma_i\otimes\sigma_i\otimes I_2+
\sigma_i\otimes\sigma_j\otimes\sigma_i\otimes\sigma_j-
\sigma_i\otimes\sigma_j\otimes\sigma_j\otimes\sigma_i)+\nonumber\\
&&\frac{\beta}{4}(\sigma_i\otimes\sigma_i\otimes I_4+
I_4\otimes\sigma_i\otimes\sigma_i+\sigma_i\otimes\sigma_i\otimes
\sigma_j\otimes\sigma_j),
\end{eqnarray}
where $i,j=1...3$ and $\sigma_i$ are spin-$\frac{1}{2}$ operators. A summation
over the repeating indices is implied in (13). 
Using the formulas (4)-(9) we see that the 
corresponding $R$-matrix has the following form:
\begin{equation}
R=\left(\begin{array}{cccccccccccccccc}
*&0&0&0&0&0&0&0&0&0&0&0&0&0&0&0\\
0&*&*&0&*&0&0&0&*&0&0&0&0&0&0&0\\
0&*&*&0&*&0&0&0&*&0&0&0&0&0&0&0\\
0&0&0&*&0&0&0&0&0&0&0&0&0&0&0&0\\
0&*&*&0&*&0&0&0&*&0&0&0&0&0&0&0\\
0&0&0&0&0&*&0&0&0&0&*&0&0&0&0&0\\
0&0&0&0&0&0&*&0&0&*&0&0&0&0&0&0\\
0&0&0&0&0&0&0&*&0&0&0&*&0&*&*&0\\
0&*&*&0&*&0&0&0&*&0&0&0&0&0&0&0\\
0&0&0&0&0&0&*&0&0&*&0&0&0&0&0&0\\
0&0&0&0&0&*&0&0&0&0&*&0&0&0&0&0\\
0&0&0&0&0&0&0&*&0&0&0&*&0&*&*&0\\
0&0&0&0&0&0&0&0&0&0&0&0&*&0&0&0\\
0&0&0&0&0&0&0&*&0&0&0&*&0&*&*&0\\
0&0&0&0&0&0&0&*&0&0&0&*&0&*&*&0\\
0&0&0&0&0&0&0&0&0&0&0&0&0&0&0&*
\end{array}\right)+O(\la^7),
\end{equation}
(here by $*$ we have marked all nonzero elements).
Up to the power $\la^6$ the matrix elements of (14) satisfy the following 
general system of relations:
\begin{equation}
R_{ij}=R_{ji},\qquad R_{ij}=R_{17-i,17-j},
\end{equation}
as well as the following one:
\begin{eqnarray}
&&R_{11}=R_{44},\quad R_{22}=R_{33}=R_{55}=R_{88}=\frac{R_{66}+R_{77}}{2},
\quad R_{25}=R_{39}=-R_{35}=-R_{29},\nonumber\\
&&R_{23}=R_{59},\quad R_{6,11}=R_{23}+R_{29}=R_{11}-R_{66},\quad
R_{7,10}=R_{23}-R_{29}=R_{11}-R_{77}.
\end{eqnarray}
So all of them may be expressed (up to $\la^6$) from $R_{11}$, $R_{66}$, and
$R_{77}$. 

According to the following relations:
\begin{equation}
\frac{R_{11}}{R_{66}}=1+(\b-\a)c(\la)+O(\la^7),\quad
\frac{R_{11}}{R_{77}}=1+(\b+\a)c(\la)+O(\la^7),
\end{equation}
where
\begin{equation}
c(\la)=\la+\frac{1}{3}(\a^2-\b^2)\la^3+
\frac{2}{15}(\a^2-\b^2)^2\la^5+O(\la^7).
\end{equation}
we may substitute together with (15) and (16) the following ansatz:
\begin{equation}
R_{11}=(1+(\b-\a)c)(1+(\a+\b)c),\quad R_{66}=1+(\b+\a)c,\quad
R_{77}=1+(\b-\a)c,
\end{equation}
and obtain from (1) the following condition on $c(\la)$:
\begin{equation}
(\a^2-\b^2)c(\la-\mu)c(\la)c(\mu)+c(\la-\mu)-c(\la)+c(\mu)=0.
\end{equation}
In the limit $\la\rightarrow\mu$ we obtain from (20) the following differential
equation:
\begin{equation}
c'(\lambda)=c'(0)(1+(\a^2-\b^2)c^2).
\end{equation}
Its solution depends on the sign of $\a^2-\b^2$ and may be taken in the form:
$c(\la)=\frac{1}{\sqrt{\a^2-\b^2}}{\rm tg}\la$ when $\a^2>\b^2$,
$c(\la)=\frac{1}{\sqrt{\b^2-\a^2}}{\rm th}\la$ when $\b^2>\a^2$, or
$c(\la)=\la$ when $\a^2=\b^2$.

\end{document}